\documentclass[fleqn,usenatbib]{mnras}

\newcommand\simlt{\lower.5ex\hbox{$\; \buildrel < \over \sim \;$}}
\newcommand\simgt{\lower.5ex\hbox{$\; \buildrel > \over \sim \;$}}

\usepackage{amsmath}
\usepackage{graphicx}	
\usepackage{amsmath}	

\title[FFE]{Breakdown of force-free electrodynamics in electric zones}
\author[Amir Levinson]{
Amir Levinson$^{1}$\thanks{E-mail: levinson@tauex.tau.ac.il}\\
$^{1}$ School of Physics and Astronomy, Tel Aviv University, Tel Aviv 69978, Israel\\}
\begin{document}
\label{firstpage}
\pagerange{\pageref{firstpage}--\pageref{lastpage}}
\maketitle

\begin{abstract}
It is shown that force-free electrodynamics (FFE) breaks down in regions where $B^2 -E^2 <0$ (electric zones)
even if $\pmb{E}\cdot\pmb{B} =0$.  Spontaneous creation of such regions will inevitably lead
to plasma oscillations that will subsequently decay over a few periods via anomalous heating and, under certain conditions, 
emission of high energy quanta, until the system relaxes to a state in which $B^2-E^2 \simlt0$. 
For M87, assuming pair plasma, the inverse Compton cooling time is 
estimated to be shorter than the dynamical time when $E^2/B^2-1 > (10^4/\sigma)^2$ roughly, where $\sigma$ is
the magnetization.  If the 
electric zone is weak, the global system will maintain a nearly force-free state, however, 
the force-free condition, $F^{\mu\nu}J_\nu=0$, will be broken  at the order of the 
excess electric field and cannot describe wave dynamics.
Our analysis does not support 
recent claims, that creation of electric zones can trigger a transition  to  force-free turbulence which, when 
generated in the ergosphere of a Kerr black hole, can lead to extraction of the black hole rotational energy.
Whether some secondary electromagnetic modes produced in the decaying electric zone can extract the BH energy 
is yet an open question.
\end{abstract}

\section{Introduction}

Astrophysical plasmas are  well described by magnetohydrodynamic theory (MHD).   
Closure of the MHD equations requires a relation between the electric 4-current 
and the remaining variables of the MHD flow, know as "generalized Ohm's law". 
A proper derivation of the generalized Ohm's law usually involves multi-fluid considerations \citep{blackman1993,gedalin1996,meier2004,most2022},
however, simplified prescriptions are commonly adopted in many situations within the framework of single-fluid MHD, most commonly 
ideal MHD.
In certain systems, notably magnetospheres of pulsars and black holes, the electromagnetic field is so 
strong that the inertia and pressure of the plasma can be ignored.  Then, closure of the electromagnetic equations 
can be obtained by adopting the force-free condition, $F^{\mu\nu}J_\nu =0$, where $F^{\mu\nu}$ is the electromagnetic 
tensor and $J^\nu$ the 4-current \citep{uchida1997a,komissarov2002,gralla2014,gralla2019}.  From this condition it readily follows that 
$\pmb{E}\cdot\pmb{B}=0$ (see below), however, the second invariant, $F\cdot F = B^2-E^2$, remains unspecified.
The reason is that since the electric 4-current $J^\mu$ does not have to be a timelike vector field, existence of a
timelike zero eigenvector of $F^{\mu\nu}$ (or a timelike generator of magnetic flux) is not guaranteed \citep{uchida1997a}.
The assumption underlying force-free electrodynamics (FFE) is that charges are sufficiently abundant 
to screen the electric field, yet exchange negligible stress-energy with the electromagnetic field. 
The conventional wisdom has been that this condition also implies $B^2-E^2 >0$,
which guarantees that the equations of FFE are hyperbolic (and, hence, are well-posed) \citep{uchida1997a,komissarov2002,pfeiffer2013,gralla2019}.

It has been argued recently \citep[][BG22]{BG2022} that the FFE equations remain valid also in 
regions where $B^2 - E^2 <0$ (termed "electric zone" in BG22),
and that in such regions "a rapid formation of fast magnetosonic (F) and Alfv\'en (A) modes ensues, with wavelengths on all scales up to the 
size of the (marginally) electric volume".  The inherent assumption underlying this argument
is that the inertia associated with the minimum charge density required 
to carry the electric 4-current (the so-called Goldriech-Julian density) is negligibly small also in such regions.
Below we refute these claims by showing that in electric zones the plasma undergoes electrostatic oscillations with a period equal
to the relativistic plasma frequency.  These oscillations, which are ignored in BG22, violate the force-free condition 
(to order $|F\cdot F|$, see further discussion below). 
It is anticipated that under the conditions prevailing in BH magnetospheres, the energy of the oscillating plasma 
will either be converted to heat via anomalous friction or, if the magnetization is high enough,
be quickly radiated away (most likely in the form of gamma-ray emission). At any rate, a spontaneously created electric zone
is likely to be screened on skin depth scale (unless forced by the global topology of the magnetic field as, e.g., in current sheets).
Contrary to the claim made in BG22, that formation of electric zones leads to turbulence via efficient generation of F and A modes,
and that the F modes thereby generated can extract the BH rotational energy, we argue that dissipation in electric zones  is governed
by plasma oscillations that will convert the excess electric field energy into heat and (non force-free) plasma turbulence.   
Most likely, dragging of magnetic surfaces into the BH will result in formation of an equatorial current sheet that 
separates field lines that cross the inner light surface and field lines that remain outside, as seen in recent 
GRMHD (e.g., \citealt{chashkina2021,ripperda2022})  and GRPIC \citep{parfrey2019,crinquand2021} simulations.

The reason why FFE breaks down in electric zones  can be traced to the dynamics of a test charge in an electromagnetic field.  Consider
the motion of a charged particle under the influence of constant magnetic and electric fields, $\pmb{B}_0$ and $\pmb{E}_0$, 
that satisfy $\pmb{E}_0\cdot\pmb{B}_0=0$.  In the case $B_0^2 > E_0^2$ a frame moving at a constant 3-velocity $\pmb{v}_B = (\pmb{E}_0\times \pmb{B}_0)/B_0^2$ exists in which the electric field vanishes \citep{jackson_classical_1999}.  In this frame, the force acting on a particle moving along the magnetic field 
direction vanishes, viz., $\pmb{v}\times\pmb{B}_0'=0$.   For an ensemble of particles moving along the magnetic field direction $\sum_i q_i (\pmb{v}_i\times\pmb{B}'_0) 
= \pmb{J}\times \pmb{B}'_0=0$, where $q_i$ is the  charge of the $i_{th}$ particle and $\pmb{J}$ the electric current. 
In covariant form it is expressed as $F'_{\mu\nu}J^\mu=0$ (since $\pmb{E}_0=0$), indicating that a force-free state can
be established in magnetic zones ($B^2>E^2$).    
On the other hand, in the case $B_0^2 < E_0^2$ the electric field cannot be transformed away. In fact,
there exists a frame moving at a constant 3-velocity 
$\pmb{v}_E = (\pmb{E}_0\times \pmb{B}_0)/E_0^2$ in which the magnetic field vanishes.   A particle of charge $q_i$ in this frame is, inevitably, 
subject to a force $q_i \pmb{E}'_0$ that accelerates it along the direction of the electric field.   Summing over an ensemble of particles 
as before yields $\pmb{J}\cdot\pmb{E} \ne 0$, or in covariant form $F'_{\mu\nu}J^\mu \ne 0$. This heuristic argument suggests that 
the FFE equations are inapplicable in electric zones, as confirmed below by detailed analysis.

A more rigorous argument (by \citealt{uchida1997a}), that applies to any spacetime, is as follows: 
When the electromagnetic tensor is electric, i.e., $F\cdot F <0$, all
zero eigenvectors of $F_{\mu\nu}$ are spacelike.  Now, although the description of FFE has no reference to the underline plasma, physically,
plasma must be present to screen the field.  Since the velocity of the charged particles, $u^\mu$,  
is a timelike vector field, it must satisfy $F_{\mu\nu}u^\nu\ne0$.  This means that all particles must cross magnetic surfaces 
and should accelerate along the electric field.   This by itself does not, necessarily, imply $F_{\mu\nu}J^\mu\ne0$, but it seems 
difficult to avoid.  A generic proof needs to show that electrically dominated force-free equilibrium is inconsistent with
the Vlasov equation and, therefore, nonphysical.  Below, we demonstrate this for a cold plasma in flat spacetime. 

From the point of view of kinetic theory, the MHD limit applies when the variations of the system are slow, more precisely,
when the characteristic frequency of the MHD waves is much smaller than the plasma and gyro frequencies.  
In magnetic regions ($B^2 - E^2 >0$) the electric conductivity along magnetic field lines 
diverges like $\omega_p^2/\omega^2$ as $\omega\to0$ (for a cold plasma with no friction, see appendix \ref{app:low_omega} for details), 
implying $\pmb{E}\cdot\pmb{B}=0$.  In the limit of high magnetization, $\sigma\gg1$, the conductivity across magnetic field lines tends 
to zero as $\sigma^{-1}$, hence the cross field current vanishes in the limit $\sigma\to\infty$, yielding the force-free condition
$F^{\mu\nu}J_\nu=0$.  In electric zones, on the other hand, particles accelerate along the electric field, implying $\pmb{J}\cdot\pmb{B}\ne0$,
so the system cannot be force-free.   More concretely, creation of electric zones always 
generates oscillations near or above the  plasma frequency, which violate the force-free
condition to the order of the excess electric field, roughly $-F\cdot F=E^2-B^2$ (see \S \ref{sec:violation}).  
We emphasize that the plasma inertia may be globally insignificant if the electric zone is small, however,
it is important at the level of wave dynamics.  In reality, even if radiative cooling 
and pair creation can be ignored, it is anticipated that the oscillations induced by the electric zone 
will damp quickly, owing to the rapid growth of instabilities 
induced by the counter oscillating beams of positive and negative charges, 
and that the bulk kinetic motion of the beams will be converted into heat and turbulence (see \S \ref{sec:dissipation}).

In what follows we analyse the dynamics of electric zones using a two-fluid approach. In appendix \ref{app:vlasov}
we derive the two-fluid equations from the Vlasov equation.   We also show there that the exact solution 
of the Vlasov equation for a uniform cold plasma coincides with that of the two fluid equations.

\section{A simplified  model for plasma dynamics}
%
Consider a  magnetized plasma consisting of positive and negative charges with 
masses $m_+$ and $m_-$ respectively (for pair plasma $m_+ = m_-$), proper densities 
$n_+$ and $n_-$ and 4-velocities $u^\mu_+$, $u^\mu_-$.   For simplicity, let us ignore, for now, pair creation,
radiative losses and other resistive effects (the inclusion of radiative losses and anomalous friction 
is considered in \S \ref{sec:dissipation}).   The particle number of each species is then conserved,
\begin{equation}
\partial_\mu (n_\pm u^\mu_\pm)=0.\label{eq:continuity}
\end{equation}
In the absence of radiative losses and energy and momentum exchange between the plasma components, energy-momentum conservation of each charged species reads
\begin{equation}
\partial_\mu T_{\pm}^{\mu\nu} = \pm e n_\pm F^{\mu\nu} u_{\pm \nu},
\label{eq:T=F}
\end{equation}
where $T_\pm^{\mu\nu}$  denote the energy momentum tensor of positive (+) and negative (-) charges,
and $F^{\mu\nu}$ is the electromagnetic tensor. It satisfies the covariant Maxwell's equations (in units with $c=1$),
\begin{equation}
\begin{aligned}
\partial_\mu F^{\nu\mu} &= 4\pi J^\nu ,\\
\partial_\mu \, ^\star F^{\mu\nu} &= 0, \label{eq:Maxwell_covariant}
\end{aligned}
\end{equation}
here $^\star F^{\mu\nu}$ is the dual electromagnetic tensor and
\begin{equation} 
J^\nu = e(n_+ u^\nu_+ - n_- u^\nu_-) \label{eq:jmu}
\end{equation}
the electric 4-current.   
Note that the total energy-momentum tensor of the plasma, $T^{\mu\nu} = T_{+}^{\mu\nu} + T_{-}^{\mu\nu}$, satisfies 
\begin{equation}
\partial_\mu T^{\mu\nu} =  F^{\mu\nu} J_\nu.
\label{eq:T=FJ}
\end{equation}
The above system of equations must be augmented by an equation of state.  
To illustrate the main point of this paper it is sufficient to invoke a cold plasma.
In that case
%
\begin{equation}
T_{\pm}^{\mu\nu} = n_\pm m_\pm u_\pm^\mu u_\pm^\nu.
\label{eq:coldT}
\end{equation}

For future purposes it is convenient to express equations (\ref{eq:continuity})-(\ref{eq:Maxwell_covariant}) and (\ref{eq:coldT})
in terms of the electric and magnetic fields components.  We then have:
\begin{eqnarray}
\nabla\cdot \pmb{E} &=& 4\pi \rho_e, \label{eq:M1} \\
\nabla\cdot \pmb{B} &=& 0,\\
 \partial_t \pmb{B} &=&  -\nabla\times \pmb{E}, \label{eq:M3}\\
\partial_t \pmb{E} &=& \nabla\times \pmb{B}- 4\pi \pmb{J},\label{eq:M4} \\
\partial_t (n_\pm \gamma_\pm) &=& -\nabla\cdot(n_\pm \pmb{u}_\pm), \\
\partial_t \pmb{u}_\pm  +  (\pmb{v}_\pm \cdot \nabla) \pmb{u}_\pm&=& \pm \frac{e}{m_\pm}(\pmb{E} + \pmb{v}_\pm \times \pmb{B}).\label{eq:kinetic}
\end{eqnarray}

%
%


\section{The Force-free limit}
The force free limit is imposed formally by setting $ \partial_\mu T^{\mu\nu} =0$ in Eq. (\ref{eq:T=FJ}), yielding
 $F_{\mu\nu} J^\mu =0$, or equivalently  $\pmb{J}\cdot\pmb{E} = 0$, $\rho_e \pmb{E}+\pmb{J}\times\pmb{B} =0$, from which 
$\pmb{E}\cdot\pmb{B} =0$ readily follows.    Of course, the force-free approximation must be consistent with the full solution 
of Eqs. (\ref{eq:M1})-(\ref{eq:kinetic}).   With some manipulations, the evolution equations of the force-free system 
take the form:
\begin{eqnarray}
 \partial_t \pmb{B} &=&  -\nabla\times \pmb{E}, \label{eq:MFF3}\\
\partial_t \pmb{E} &=& \nabla\times \pmb{B} \label{eq:MFF4} \\ \nonumber 
&-& \frac{\nabla \cdot \pmb{E} (\pmb{E}\times \pmb{B})+(\pmb{B}\cdot \nabla\times \pmb{B}-\pmb{E}\cdot \nabla\times \pmb{E})\pmb{B}}{B^2},   
\end{eqnarray}
subject to the initial conditions $\nabla\cdot \pmb{E} = 4\pi \rho_e$ and $\nabla\cdot \pmb{B} =0$.\\

To elucidate the wave characteristics of FFE, consider linear perturbations 
on a background of orthogonal magnetic and electric fields, $\pmb{B}_0 =B_0 \hat{n} $,  $\pmb{E}_0 = \eta\, B_0 \hat{t}$, $\hat{n}\cdot\hat{t}=0$.
Here $\eta=E_0/B_0$ measures whether the background force-free equilibrium is magnetically dominated ($\eta<1$) or electrically dominated ($\eta>1$).
For one-dimensional periodic waves, $\propto e^{i(\omega t -kz)}$, the linearized equations yield the following dispersion 
relation, expressed here in terms of the phase velocity $v=\omega/k$ (see appendix \ref{app:waves} for details):
\begin{equation}
 (1-v^2)[(v+ \eta n_1 t_2)^2 - n_3^2(1-\eta^2)] v^2 =0.  \label{eq:dispersion}
\end{equation} 
%
This dispersion equation  admits four wave modes: two fast modes with phase velocities $v_f=\pm 1$, and 
%
two additional modes with
\begin{equation}
v_a = -\eta n_1 t_2 \pm \sqrt{n_3^2(1-\eta^2)} = \frac{-B_{01}E_{02}\pm \sqrt{B_{03}^2(B_0^2-E_0^2)}}{B_0^2},
\label{eq:phase_A}
\end{equation}
consistent with the result derived in \citet{komissarov2002}. 
For $\eta <1$ (i.e., $B_0^2-E_0^2 >0$) the phase velocities $v_a$ are real and the modes correspond to Alfv\'en waves,
indicating that the  FFE system is indeed hyperbolic \citep{uchida1997b,komissarov2002}.

On the other hand, for $\eta>1$ and $n_3 \neq 0$ (i.e., $\pmb{k}\cdot\pmb{B}_0\ne0$) the phase velocities are complex, 
and it appears that there is a growing mode for each choice
of a $k$ vector.  The presumption that FFE is valid in this regime implies that the energy of the growing mode cannot come from the plasma 
(as also implied by the force-free condition $\pmb{J}\cdot\pmb{E} =0$).  To the understanding of this author, this is the basis for the
claim made in BG22, that electric regions should lead to force-free turbulence. 
 Below it is shown that the complex solutions of Eq. (\ref{eq:phase_A}) obtained for $\eta >1$ stem from an inconsistency
 between the presumed force-free condition and the full solution of the plasma dynamics equations, Eqs. (\ref{eq:M1})-(\ref{eq:kinetic}).
%

\section{The validity of FFE}
\label{sec:violation}
We now demonstrate that FFE is valid in magnetic zones when the changes of the plasma parameters are slow, 
owing to a strong confinement by the background magnetic field, but not in electric zones under any circumstances.  
For simplicity, we suppose that initially 
the electric and magnetic fields are uniform and orthogonal, 
$\pmb{B}_0 =B_0 \hat{n} $,  $\pmb{E}_0 = \eta\, B_0 \hat{t}$, $\hat{n}\cdot\hat{t}=0$,
 and consider the subsequent evolution of the system. 
 
\subsection{The case $B_0 > E_0$}
In this case a Lorentz transformation can be performed to a frame $K'$ moving at a 3-velocity $\pmb{v}_B = (\pmb{E}_0\times\pmb{B}_0)/B_0^2$,
in which the electric field vanishes, $\pmb{E}'_0=0$, and the magnetic field is given by 
\begin{equation}
\pmb{B}'_0 = (1-E_0^2/B_0^2)^{1/2} \pmb{B}_0 = (B_0^2-E_0^2)^{1/2} \hat{n}.
\end{equation}
Since the frames $K$ and $K'$ are related through a proper Lorentz transformation, it is sufficient to 
compute the dispersion relations in $K'$. 

Wave solutions for cold magnetized plasma are derived in many textbooks.  
It is well known that in the low frequency limit the plasma waves converge to MHD modes.  
Here we elucidate the emergence of FFE in highly magnetized plasma, in order to compare with the 
analysis of electric zones outlined in the succeeding section.  In what follows we provide 
a succinct derivation.
The full analysis is outlined in appendix \ref{app:low_omega}.  

The acceleration of particles driven by a periodic wave having frequency $\omega$ and amplitude $e_0$ can be 
computed from  Eq. (\ref{eq:kinetic}).  For the parallel component
(along $\pmb{B}_0'$) we obtain
\begin{equation}
   i\omega \pmb{v}_{\pm||} = \frac{e}{m_\pm} \pmb{e}_{||},
\end{equation}
and for the perpendicular component
\begin{equation}
   i\omega \pmb{v}_{\pm\perp} = \frac{e \omega^2}{m_\pm(\omega^2-\omega_{c\pm}^2)}
    \left( \pm\pmb{e}_\perp +\frac{e}{i\omega m_\pm} \pmb{e}_\perp\times\pmb{B}'_0\right),
    \label{eq:acc_perp_wave}
\end{equation}
where $\pmb{e}_{||}$ ($\pmb{e}_\perp$) denote the wave electric field component parallel (perpendicular) to the 
background magnetic field $\pmb{B}_0'$, with normalization $\pmb{e}_{||}^2+\pmb{e}_\perp^2 = e_0^2$.  
The second term on the right hand side of Eq. (\ref{eq:acc_perp_wave})
is associated with drift motion and is independent of the sign of the charge. Its contribution to the electric current 
vanishes identically for symmetric (pair) 
plasma, and is negligibly small for electron ion plasma in the limit $\omega\to 0$ (see Eq. \ref{eq:app_low_omega}).
For the Alfv\'en mode the low frequency limit yields (see appendix \ref{app:low_omega})
 $|\pmb{e}_{||}|/|\pmb{e}_\perp| \approx (\omega/\omega_p)^2 \tan\theta $, where $\omega_p^2 =\omega_{p+}^2+\omega_{p-}^2$\footnote{This relation holds at angles $\theta$ for which $\tan\theta \ll (\omega/\omega_p)^2$.}.  Consequently, 
$|\pmb{v}_{\pm ||}|/|\pmb{v}_{\pm\perp}| \sim (\omega_{c\pm}/\omega_{p})^2 = \sigma_\pm (\omega_{p\pm}/\omega_p)^2$, where 
$\sigma_\pm$ is the magnetization. The electric current components 
satisfy $J_{\perp}/J_{||}\approx \tan\theta/\sigma$, where $\sigma^{-1} = \sigma_+^{-1}+\sigma_-^{-1}$ (see Eq. \ref{eq:J_app_B}).
We see that in the limit $\sigma_\pm \to\infty$ the 
particles are confined to move along the background magnetic field $\pmb{B}_0'$, and $J_\perp =0$.  The reason is that in this limit the inertia 
of the plasma becomes  vanishingly small.  It is also seen that in practice what drives the oscillations of the electric
current along the background magnetic field is the tiny electric field 
component $\pmb{e}_{||} \approx (\omega/\omega_p)^2\tan\theta \, e_0$, that formally vanishes as $\omega\to 0$.
This tiny component is sufficient to generate a finite electric current because the conductivity along $\pmb{B}_0'$
diverges as $\omega_p^2/\omega^2$ (see Eq. \ref{eq:conductivity_low_omega}), as also found in \citep{most2022}.  To linear order the force acting on the plasma 
is given by $\pmb{J}\times\pmb{B}'_0 \sim (i\omega/4\pi) B_0/\sigma$, indicating that 
the deviation from the force-free condition is of order $\sigma^{-1}$.

\subsection{The case $B_0 < E_0$}
\label{sec:violation_E}
In this case a Lorentz transformation can be performed to a frame $K'$ moving at a 3-velocity $\pmb{v}_E = (\pmb{E}_0\times\pmb{B}_0)/E_0^2$,
in which the magnetic field vanishes, $\pmb{B}'_0=0$, and the electric field is given by 
\begin{equation}
\pmb{E}'_0 = (1-B_0^2/E_0^2)^{1/2} \pmb{E}_0 = (E_0^2-B_0^2)^{1/2} \hat{t}.
\end{equation} 
From the equation of motion (\ref{eq:kinetic}) it is evident that perturbative solutions around the constant field $\pmb{E}'_0$
do not exist (since there is no solution with $\pmb{E}'_0 =$ const), unless $E'_0$ itself is small.  However, perturbations around a uniform time dependent solution can be sought in general (see appendix \ref{app:linear_waves}). 

\begin{figure}
  \centering
  \includegraphics[width=0.5\textwidth]{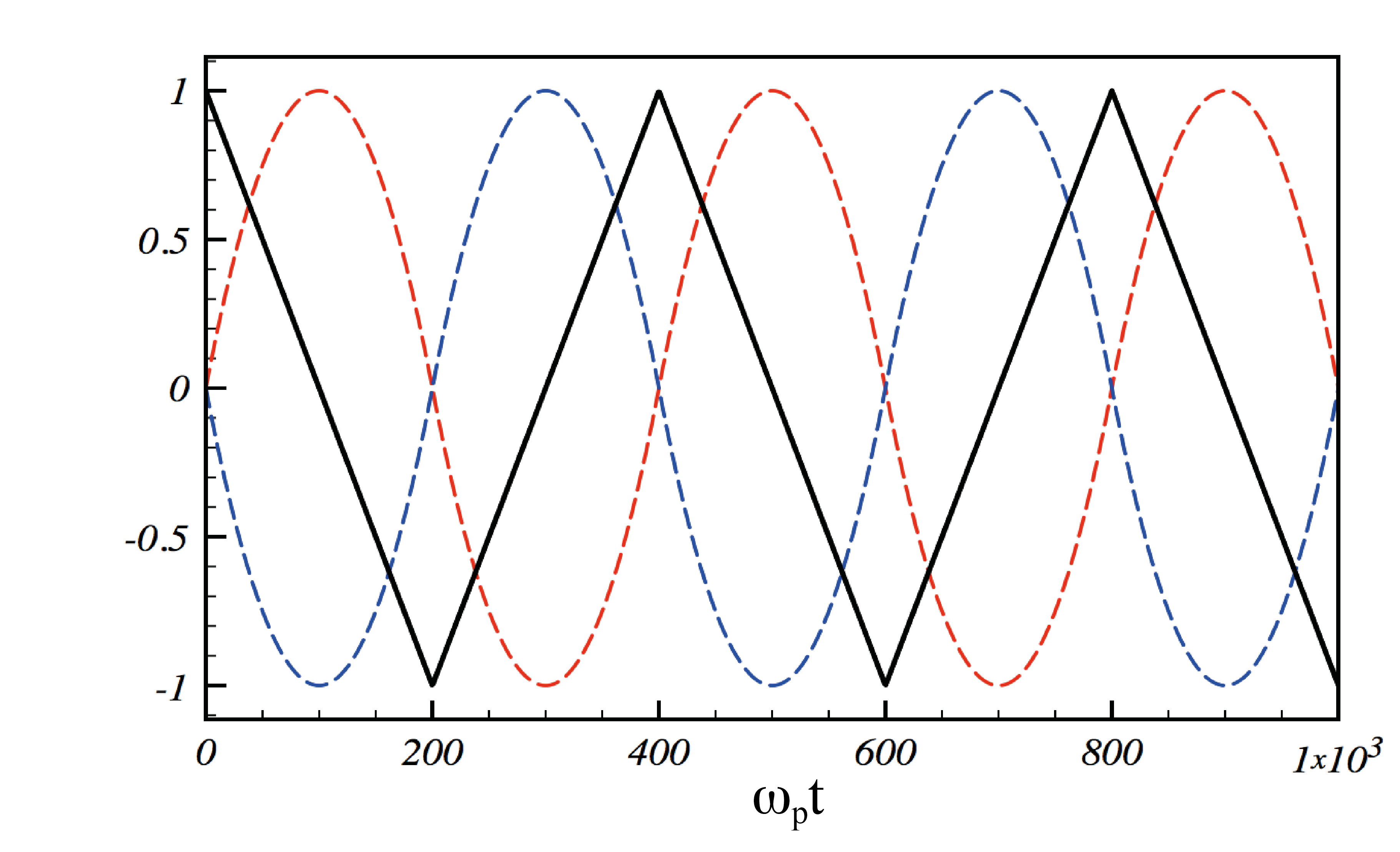}
  \vspace{-12pt}
  \caption{Periodic oscillations of the electric zone for symmetric plasma with initial electric field $E'_0 = 10^2 \sqrt{8\pi m N_0}$. 
  Shown are $E/E'_0$ (black line), $u_+/u_{max}$ (red line) and $u_-/u_{max}$ (blue line). The time $t$ is normalized by the inverse of the
  non-relativistic plasma frequency, $\omega_p = \sqrt{8\pi e^2 N_0/m}$. The apparent triangular shape of the electric field oscillations is 
  due to the relativistic motion of the plasma.}
\label{fig:ocl}
\end{figure}

Consider first a time dependent, uniform system, $\pmb{E}=\pmb{E}(t)$, $\pmb{B}=\pmb{B}(t)$.   Eq. (\ref{eq:M3}) implies that $\pmb{B}$ is independent of time, thus $\pmb{B}=\pmb{B}'_0=0$.   From the continuity equation, $\partial_t(n_\pm\gamma_\pm)=0$, we obtain $\gamma_+n_+=\gamma_- n_- =N_0$ for a neutral plasma, and $\pmb{J} = e (n_+ \pmb{u}_+ -n_-\pmb{u}_-) = eN_0 (\pmb{\beta}_+ - \pmb{\beta}_-)$ for the electric current.  
The remaining equations reduce to
\begin{eqnarray}
\partial_t \pmb{E} &=& - 4\pi \pmb{J} = - 4\pi eN_0 (\pmb{\beta}_+ -\pmb{\beta}_-),\label{eq:ME4} \\
\partial_t \pmb{u}_\pm &=& \pm \frac{e}{m_\pm}\pmb{E} .\label{eq:kineticE}
\end{eqnarray}
As seen, the system undergoes strong plasma oscillations, 
with $\pmb{J}\cdot\pmb{E} = - \partial_t (E^2/8\pi) \ne0$ and, hence,  is not force-free.  
 In fact, the energy oscillates between potential energy of the electric field and kinetic energy of the 
 plasma (Fig. \ref{fig:ocl}).
 The amplitude of the electric field oscillations is the initial field $E'_0$ and the period (in the relativistic regime) 
 can be readily shown to be $T' =  E_0'/2\pi e N_0$, using energy conservation (from Eq. (\ref{eq:T=FJ})):
 \begin{equation}
     N_0[m_+(\gamma_+-1)+ m_-(\gamma_- -1)] +\frac{E^2}{8\pi} = \frac{E_0^{'2}}{8\pi}.
     \label{eq:Ezone_cons}
 \end{equation}
 For symmetric plasma, $m_+ = m_- =m$, $\gamma_+=\gamma_- =\gamma$, the amplitude of the Lorentz factor is
  \begin{equation}
    \gamma_{max} = 1 + \frac{E_0^{'2}}{16\pi m N_0} = 1+ \frac{\eta (\eta^2-1)^{1/2}}{2}\sigma ,
    \label{eq:gamma_max}
 \end{equation}
 here $\sigma = B_0^2/4\pi(2m\gamma_E N_0)$ is the magnetization of the plasma in the original frame, in which the 
 total density is $2N_0 \gamma_E$, and $\gamma_E = (1-B_0^2/E_0^2)^{-1/2}$ is the Lorentz factor of the 
 frame $K'$ with respect to the frame in which $B_0$ is measured.  We see that the motion of the plasma is relativistic as long as $\eta^2-1 > (2/\sigma)^{2}$ roughly.
 
 If the electric zone is small, such that the plasma dynamics is non-relativistic ($\gamma_{max}\sim 1$), then 
 linear wave solutions can be sought.  The dispersion relations reduce to those of  waves in cold unmagnetized plasma.
 There are two linear modes; one electrostatic with $\omega=\omega_p$ (which is the nonrelativistic solution of 
 Eqs. \ref{eq:ME4}-\ref{eq:kineticE}),  and one transverse with $\omega^2 = k^2 +\omega_p^2$.  Transforming back to the original frame 
 it is seen that these dispersion relations are vastly different
 than those derived in Eq. (\ref{eq:phase_A}) for $\eta>1$ under the assumption that FFE is valid to all orders in this regime. 
 In particular, the frequency is real and there is no exponential instability; the complex solutions of Eq. (\ref{eq:phase_A})
 are artefacts of the improper Ohm's law invoked there.
 At any rate, $\pmb{J}\cdot\pmb{E} \approx (\omega_p^2/4\pi i\omega)E_0^{'2}$, indicating that FFE is broken at the 
 order $-F\cdot F$.  If the electric zone is small, $\eta^2-1 \ll 1$, the system will remain, to a good 
 approximation, in a force-free state ($\sigma \gg 1$ in the original frame), however, the dynamics of 
 waves generated by the creation of the electric zone cannot be described by the FFE equations.

\subsection{Dissipation of the electric zone}
\label{sec:dissipation}
The fate of the oscillations triggered by creation of an electric zone depend on the conditions in the source and the 
magnitude of the excess electric field.  For illustration, we consider here the inner magnetosphere of a rapidly rotating, supermassive 
black hole of mass $M=10^9 M_9 M_\odot$.  Curvature emission is likely unimportant since the particles do not
move along magnetic field lines.  To estimate the cooling time due to inverse Compton emission, $t_{IC}$, we invoke a soft 
radiation source (the outer disk) of luminosity $L=10^{42} L_{42}$ erg s$^{-1}$ 
and radius $R = r_g \tilde{R} = 1.5\times10^{14} M_9 \tilde{R}$ cm.
The period in the original frame, $T=\gamma_E T'$,
can be expressed in terms of the angular velocity of magnetic field lines, $\Omega\approx \omega_H/2 = c/4r_g$, and the multiplicity $\kappa = n/n_{GJ}$, where $n_{GJ}$ is the Goldreich-Julian density, as
$T \approx \sqrt{\eta^2-1}/\kappa \Omega \approx 2\times 10^{4} \eta M_9/\kappa$ s.
For the maximum Lorentz factor in Eq. (\ref{eq:gamma_max}) we then obtain 
\begin{equation}
     \frac{t_{IC}}{T} \approx 15 \frac{M_9 \tilde{R}^2}{L_{42}}\frac{\kappa} {\eta (\eta^2-1)}\frac{1}{\sigma}.
\end{equation}
Thus, the energy will be radiated away  over time shorter than the wave period provided $\eta(\eta^2-1)\sigma  > 15 \kappa M_9\tilde{R}^2/L_{42}$.
For M87 $M_9\approx 6$, $L_{42}\sim1$, $\tilde{R}\sim 5$, $B_0\sim 10^2$ G, so $(\eta^2-1) \simgt 2\times 10^3\kappa/\sigma \approx 10^{-11} \kappa^2$ is required to radiate efficiently.  Note that to radiate over time shorter that the dynamical time, that is, $\Omega t_{IC} <1$,
requires $(\eta^2-1)^{1/2} > 10^{-11} \kappa \approx 2\times10^3/\sigma$. So for $\kappa <10^{10}$, or equivalently $\sigma >10^4$ even a weak electric zone will
decay radiatively. 

\begin{figure}
  \centering
  \includegraphics[width=0.5\textwidth]{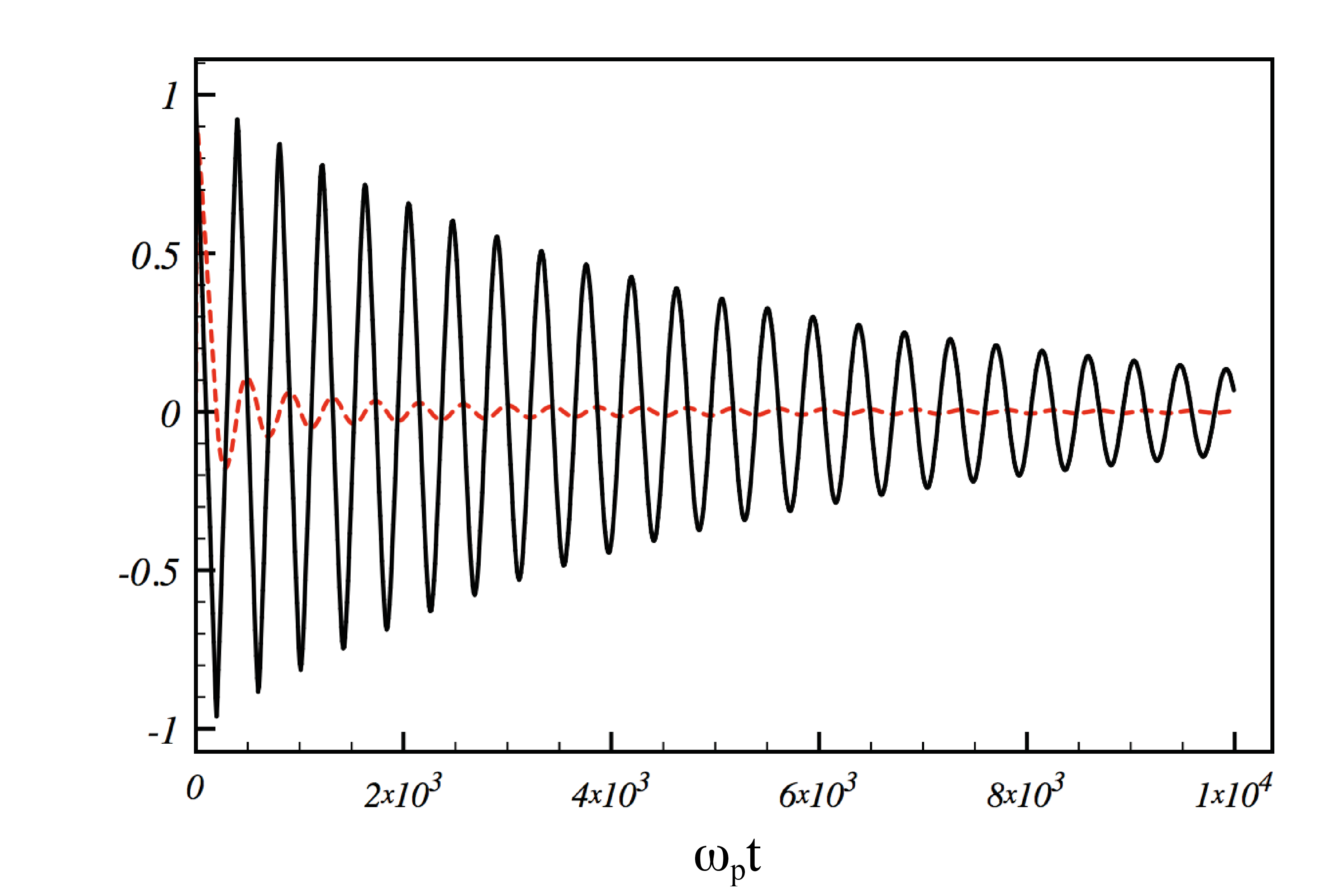}
  \vspace{-12pt}
  \caption{Decaying oscillations in a model with friction.  Shown are $\tilde{E}/\tilde{E}_0'$ (black) and $u_+/u_{max}$ (red).}
\label{fig:damp}
\end{figure}

If the cooling time is considerably longer than the period of plasma oscillations, it is likely that the 
energy of the oscillating plasma will be converted into heat via anomalous friction. 
Ab initio calculations of such damping requires kinetic plasma simulations with sufficint resolution, which
are beyond the scope of this paper.  Below we adopt the phenomenological prescription described in \cite{levinson2020}
to model internal friction. In this treatment the friction force is given by
\begin{equation}
g^\mu_\pm =  \pm \chi m_- c n_+ n_- (u^\mu_- - u^\mu_+),     
\end{equation}
where $\chi$ is a dynamical coefficient, assumed to be constant. This force is added to the right hand side of Eq. (\ref{eq:T=F}).
Since the plasma is heated as a result of this friction, we must include the specific enthalpy $h_\pm$ in the energy momentum tensors
$T_\pm^{\mu\nu}$.  For simplicity, we suppose that $h_+ = h_- =h$ (which is expected in general).
Introducing the dimensionless quantities $\tau = \omega_p t$, $\tilde{E} = E/\sqrt{8\pi m N_0}$
and $\tilde{\chi} = \chi N_0/\omega_p$, where $\omega_p = \sqrt{8\pi N_0 e^2/m}$, we arrive at the set of equations
\begin{eqnarray}
\partial_\tau \tilde{E}  =  - \frac{1}{2} (\beta_+ - \beta_-),\label{eq:MH4} \\
\partial_\tau (h u_\pm) = \pm \tilde{E} \pm \frac{\tilde{\chi}}{\gamma_+\gamma_-}(u_- -u_+),\label{eq:kineticH}\\
  h \gamma_+ + h \gamma_-  +\tilde{E}^2 = \tilde{E}_0^{'2} +2.
\end{eqnarray}
A solution is shown in Fig. \ref{fig:damp} for $\tilde{\chi}=1$ and the initial condition $u_\pm=0, h=1, \tilde{E}=\tilde{E}'_0=10^2$.
From the above equations we find a decay time of $\tau_{decay} \approx \tilde{E}_0^{'2}/8\tilde{\chi}$ in the frame $K'$.
In the original frame, the decay time is $t_{decay}= \gamma_E \omega_p^{-1}\tau_{decay}$. In terms of the angular frequency of field lines, 
$\Omega$, and the multiplicity, $\kappa$, the decay time in the BH frame can be expressed as:
\begin{equation}
    \Omega \, t_{decay} \approx  \frac{\eta^{5/2}}{32(\eta^2-1)^{1/4}}\frac{\sqrt{\sigma}}{\tilde{\chi}\kappa}.
    \label{eq:decay}
\end{equation}
The oscillations will decay over time shorter than the dynamical time 
provided $\eta^{-5/2}(\eta^2-1)^{1/4} > \sqrt{\sigma}/(32\tilde{\chi}\kappa)$, or in case of M87, 
$\eta^{-5/2}(\eta^2-1)^{1/4} > 3\times10^5/\tilde{\chi}\kappa^{3/2}$.
However, $E_0'$ must be large enough to
generate the instability that couples the two beams.

We emphasize that this simplified model is probably unrealistic, as it ignores the effect of plasma turbulence, anticipated to be generated by 
the counter-propagating beam instability, on the evolution of the system.  In particular, the decaying electric field is not expected to 
exhibit the coherent oscillations seen in Fig. \ref{fig:damp} for times longer than the growth time of the instability.  Nonetheless, 
it demonstrates that the field should decay and that the major fraction of its energy will most likely turn into random kinetic energy.
A fraction of the decaying E-zone energy might escape the E-zone in the form of FFE waves.  What fraction, and how these waves 
are transmitted to the magnetic zone is yet an open issue.  Some insights might be provided by future PIC simulations with appropriate  setup.



\section{Conclusions}
The main conclusion is that force-free electrodynamics is valid only in regions where $B^2 - E^2 >0$.
Spontaneous creation of electric zones ($B^2-E^2 <0$) triggers plasma oscillations  that transfer the energy of
the excess electric field to the plasma. 
Under conditions anticipated in the magnetospheres of compact astrophysical objects these oscillations are expected to decay 
quickly (over a few  periods) via
generation of kinetic turbulence and heat, and/or emission of high-energy photons if the magnetization is high enough.  
The created electric zone will be subsequently 
screened and the system will relax into a state where $B^2 \sim E^2$.

The analysis in sections \ref{sec:violation_E} and \ref{sec:dissipation} assumes instantaneous creation of 
electrically dominated region.  In reality, it is expected that formation of the electric zone will occur 
gradually over the light crossing time of the electric volume.   Equation (\ref{eq:decay}) then suggests that dissipation by anomalous 
friction will prevent the growth of the excess electric field beyond the instability threshold, leading to 
adiabatic evolution of the system that  converts the energy of the excess field to heat.  Self-consistent calculations require 
kinetic plasma simulations.

A particular example, studied recently using kinetic plasma simulations \citep{Li2021}, is collision of 
nonlinear Alfv\'en waves.  When the amplitude of the colliding waves is large enough, an electric zone forms
in the overlapping region of the waves.  The simulations confirm that the collision triggers rapid dissipation 
via particle acceleration, and that the electric field generated in the 
overlapping zone is indeed screened. 

 If E-zones in the configuration envisaged by BG22 (rather than equatorial current sheet) indeed form in black hole magnetospheres,
the question arises as to why they are not seen in current GRPIC simulations  \citep{parfrey2019,crinquand2021}, that resolve the skin 
depth and are able to identify electrostatic oscillations in starved magnetospheric regions (in which $\pmb{E}\cdot{\pmb B}\ne0$).  
All current simulations that invoke physical magnetic field configuration (monopole field excluded) exhibit formation of a current sheet, 
which to this author seems natural by virtue of the fact that field lines crossing the fast magnetosonic surface cannot be pushed 
backwards by lack of causality.  Nonetheless, it is desirable  to think of setups that might be more suitable to test the hypothesis of BG22.

\section*{acknowledgment}
I thank Roger Blandford, Sasha Phillipov and Arno Vanthieghem for enlightening discussions, and Sam Gralla and Jens Mahlmann for useful comments. 

\section*{Data Availability}
There are no new data associated with this article.	

\bibliographystyle{mnras}
\bibliography{FFE}

\begin{thebibliography}{}
\makeatletter
\relax
\def\mn@urlcharsother{\let\do\@makeother \do\$\do\&\do\#\do\^\do\_\do\%\do\~}
\def\mn@doi{\begingroup\mn@urlcharsother \@ifnextchar [ {\mn@doi@}
  {\mn@doi@[]}}
\def\mn@doi@[#1]#2{\def\@tempa{#1}\ifx\@tempa\@empty \href
  {http://dx.doi.org/#2} {doi:#2}\else \href {http://dx.doi.org/#2} {#1}\fi
  \endgroup}
\def\mn@eprint#1#2{\mn@eprint@#1:#2::\@nil}
\def\mn@eprint@arXiv#1{\href {http://arxiv.org/abs/#1} {{\tt arXiv:#1}}}
\def\mn@eprint@dblp#1{\href {http://dblp.uni-trier.de/rec/bibtex/#1.xml}
  {dblp:#1}}
\def\mn@eprint@#1:#2:#3:#4\@nil{\def\@tempa {#1}\def\@tempb {#2}\def\@tempc
  {#3}\ifx \@tempc \@empty \let \@tempc \@tempb \let \@tempb \@tempa \fi \ifx
  \@tempb \@empty \def\@tempb {arXiv}\fi \@ifundefined
  {mn@eprint@\@tempb}{\@tempb:\@tempc}{\expandafter \expandafter \csname
  mn@eprint@\@tempb\endcsname \expandafter{\@tempc}}}

\bibitem[\protect\citeauthoryear{{Blackman} \& {Field}}{{Blackman} \&
  {Field}}{1993}]{blackman1993}
{Blackman} E.~G.,  {Field} G.~B.,  1993, \mn@doi [\prl]
  {10.1103/PhysRevLett.71.3481}, \href
  {https://ui.adsabs.harvard.edu/abs/1993PhRvL..71.3481B} {71, 3481}

\bibitem[\protect\citeauthoryear{{Blandford} \& {Globus}}{{Blandford} \&
  {Globus}}{2022}]{BG2022}
{Blandford} R.,  {Globus} N.,  2022, \mn@doi [\mnras] {10.1093/mnras/stac1682},
  \href {https://ui.adsabs.harvard.edu/abs/2022MNRAS.tmp.1627B} {}

\bibitem[\protect\citeauthoryear{{Chashkina}, {Bromberg}  \&
  {Levinson}}{{Chashkina} et~al.}{2021}]{chashkina2021}
{Chashkina} A.,  {Bromberg} O.,   {Levinson} A.,  2021, \mn@doi [\mnras]
  {10.1093/mnras/stab2513}, \href
  {https://ui.adsabs.harvard.edu/abs/2021MNRAS.508.1241C} {508, 1241}

\bibitem[\protect\citeauthoryear{{Crinquand}, {Cerutti}, {Dubus}, {Parfrey}  \&
  {Philippov}}{{Crinquand} et~al.}{2021}]{crinquand2021}
{Crinquand} B.,  {Cerutti} B.,  {Dubus} G.,  {Parfrey} K.,   {Philippov} A.,
  2021, \mn@doi [\aap] {10.1051/0004-6361/202040158}, \href
  {https://ui.adsabs.harvard.edu/abs/2021A&A...650A.163C} {650, A163}

\bibitem[\protect\citeauthoryear{{Gedalin}}{{Gedalin}}{1996}]{gedalin1996}
{Gedalin} M.,  1996, \mn@doi [\prl] {10.1103/PhysRevLett.76.3340}, \href
  {https://ui.adsabs.harvard.edu/abs/1996PhRvL..76.3340G} {76, 3340}

\bibitem[\protect\citeauthoryear{{Gralla} \& {Iqbal}}{{Gralla} \&
  {Iqbal}}{2019}]{gralla2019}
{Gralla} S.~E.,  {Iqbal} N.,  2019, \mn@doi [\prd]
  {10.1103/PhysRevD.99.105004}, \href
  {https://ui.adsabs.harvard.edu/abs/2019PhRvD..99j5004G} {99, 105004}

\bibitem[\protect\citeauthoryear{{Gralla} \& {Jacobson}}{{Gralla} \&
  {Jacobson}}{2014}]{gralla2014}
{Gralla} S.~E.,  {Jacobson} T.,  2014, \mn@doi [\mnras]
  {10.1093/mnras/stu1690}, \href
  {https://ui.adsabs.harvard.edu/abs/2014MNRAS.445.2500G} {445, 2500}

\bibitem[\protect\citeauthoryear{Jackson}{Jackson}{1999}]{jackson_classical_1999}
Jackson J.~D.,  1999, Classical electrodynamics, 3rd ed. edn.
Wiley, New York, {NY}, \url {http://cdsweb.cern.ch/record/490457}

\bibitem[\protect\citeauthoryear{{Komissarov}}{{Komissarov}}{2002}]{komissarov2002}
{Komissarov} S.~S.,  2002, \mn@doi [\mnras] {10.1046/j.1365-8711.2002.05313.x},
  \href {https://ui.adsabs.harvard.edu/abs/2002MNRAS.336..759K} {336, 759}

\bibitem[\protect\citeauthoryear{{Levinson}}{{Levinson}}{2020}]{levinson2020}
{Levinson} A.,  2020, \mn@doi [\pre] {10.1103/PhysRevE.102.063210}, \href
  {https://ui.adsabs.harvard.edu/abs/2020PhRvE.102f3210L} {102, 063210}

\bibitem[\protect\citeauthoryear{{Li}, {Beloborodov}  \& {Sironi}}{{Li}
  et~al.}{2021}]{Li2021}
{Li} X.,  {Beloborodov} A.~M.,   {Sironi} L.,  2021, \mn@doi [\apj]
  {10.3847/1538-4357/abfe5f}, \href
  {https://ui.adsabs.harvard.edu/abs/2021ApJ...915..101L} {915, 101}

\bibitem[\protect\citeauthoryear{{Meier}}{{Meier}}{2004}]{meier2004}
{Meier} D.~L.,  2004, \mn@doi [\apj] {10.1086/382201}, \href
  {https://ui.adsabs.harvard.edu/abs/2004ApJ...605..340M} {605, 340}

\bibitem[\protect\citeauthoryear{{Parfrey}, {Philippov}  \&
  {Cerutti}}{{Parfrey} et~al.}{2019}]{parfrey2019}
{Parfrey} K.,  {Philippov} A.,   {Cerutti} B.,  2019, \mn@doi [\prl]
  {10.1103/PhysRevLett.122.035101}, \href
  {https://ui.adsabs.harvard.edu/abs/2019PhRvL.122c5101P} {122, 035101}

\bibitem[\protect\citeauthoryear{{Pfeiffer} \& {MacFadyen}}{{Pfeiffer} \&
  {MacFadyen}}{2013}]{pfeiffer2013}
{Pfeiffer} H.~P.,  {MacFadyen} A.~I.,  2013, arXiv e-prints, \href
  {https://ui.adsabs.harvard.edu/abs/2013arXiv1307.7782P} {p. arXiv:1307.7782}

\bibitem[\protect\citeauthoryear{{Rendall}}{{Rendall}}{1996}]{rendall199}
{Rendall} A.~D.,  1996, arXiv e-prints, \href
  {https://ui.adsabs.harvard.edu/abs/1996gr.qc.....4001R} {pp gr--qc/9604001}

\bibitem[\protect\citeauthoryear{{Ripperda}, {Liska}, {Chatterjee}, {Musoke},
  {Philippov}, {Markoff}, {Tchekhovskoy}  \& {Younsi}}{{Ripperda}
  et~al.}{2022}]{ripperda2022}
{Ripperda} B.,  {Liska} M.,  {Chatterjee} K.,  {Musoke} G.,  {Philippov} A.~A.,
   {Markoff} S.~B.,  {Tchekhovskoy} A.,   {Younsi} Z.,  2022, \mn@doi [\apjl]
  {10.3847/2041-8213/ac46a1}, \href
  {https://ui.adsabs.harvard.edu/abs/2022ApJ...924L..32R} {924, L32}

\bibitem[\protect\citeauthoryear{{Uchida}}{{Uchida}}{1997a}]{uchida1997a}
{Uchida} T.,  1997a, \mn@doi [\pre] {10.1103/PhysRevE.56.2198}, \href
  {https://ui.adsabs.harvard.edu/abs/1997PhRvE..56.2198U} {56, 2198}

\bibitem[\protect\citeauthoryear{{Uchida}}{{Uchida}}{1997b}]{uchida1997b}
{Uchida} T.,  1997b, \mn@doi [\mnras] {10.1093/mnras/291.1.125}, \href
  {https://ui.adsabs.harvard.edu/abs/1997MNRAS.291..125U} {291, 125}

\bibitem[\protect\citeauthoryear{{Van Putten}, {Levinson}  \& {t'Hooft}}{{Van
  Putten} et~al.}{2012}]{lev2012}
{Van Putten} M. H.~P.~M.,  {Levinson} A.,   {t'Hooft} F. b.~G.,  2012,
  {Relativistic Astrophysics of the Transient Universe}

\makeatother
\end{thebibliography}

\appendix 
\section{Linear force-free waves}\label{app:waves}

Consider  linear perturbations of the force free equations (\ref{eq:MFF3})-(\ref{eq:MFF4})
on the background magnetic and electric fields, $\pmb{B}_0 =B_0 \hat{n} $,  $\pmb{E}_0 = \eta\, B_0 \hat{t}$, $\hat{n}\cdot\hat{t}=0$, 
and denote the wave fields as $\pmb{b}(t,\pmb{r})$, $\pmb{e}(t,\pmb{r})$ (where $\pmb{B}(t,\pmb{r}) =  \pmb{B}_0+  \pmb{b}(t,\pmb{r})$ and $ \pmb{E}(t,\pmb{r}) =  \pmb{E}_0+  \pmb{e}(t,\pmb{r})$).
The electric current is expanded as $ \pmb{J} =  \pmb{J}_0 + \delta\pmb{J}$, where for the uniform background field $ \pmb{J}_0 =0$.
The linearized equations read:
\begin{equation}
\begin{aligned}
\frac{1}{c}\partial_t\pmb{b} &= -\nabla\times \pmb{e},\\
\frac{1}{c}\partial_t\pmb{e} &= \nabla\times \pmb{b} \\
&-  \frac{\nabla\cdot\pmb{e} (\pmb{E}_0\times \pmb{B}_0)+(\pmb{B}_0\cdot \nabla\times \pmb{b}-\pmb{E}_0\cdot \nabla\times \pmb{e})\pmb{B}_0}{B_0^2}.
\end{aligned}
\end{equation}
In the short wavelength approximation, $\pmb{e},\pmb{b} \propto e^{i(\omega t-\pmb{k}\cdot\pmb{r)}}$, we have:
\begin{equation}
\begin{aligned}
\omega \pmb{b} &= \pmb{k}\times \pmb{e},\\
\omega \pmb{e} &= -\pmb{k}\times \pmb{b} + \eta(\pmb{k}\cdot\pmb{e}) (\hat{t}\times \hat{n}) \\
&+    \hat{n}\cdot(\pmb{k}\times \pmb{b})\hat{n} - \eta \, \hat{t}\cdot(\pmb{k}\times \pmb{e})\hat{n},
\end{aligned}
\end{equation}
from which we obtain
\begin{equation}
\begin{aligned}
\omega^2 \pmb{e} &= k^2 \pmb{e} -(\pmb{k}\cdot\pmb{e})\pmb{k} +\omega \eta(\pmb{k}\cdot\pmb{e}) (\hat{t}\times \hat{n}) \\
&+ [(\pmb{k}\cdot\pmb{e})\pmb{n}\cdot \pmb{k} 
- k^2 \pmb{n}\cdot\pmb{e} -\omega \eta \, \hat{t}\cdot(\pmb{k}\times \pmb{e})]\pmb{n}.
\label{eq:dispersion_1}
\end{aligned}
\end{equation}
For 1D waves propagating in the $z$-direction, $\pmb{k} = k \hat{z}$.  Without any loss of generality we can choose $\pmb{B}_0$ to lie in the plane $(x,z)$, so that $n_2=0$.  Then 
$n_3^2 = 1 -n_1^2$, and $(\hat{t}\times \hat{n})_2 = -t_1/n_3$.  Equation (\ref{eq:dispersion_1}) can be expressed as 
\begin{equation}
\Lambda_{ij} e_j =0,
\end{equation}
where the matrix $\Lambda$ is given explicitly by
\begin{equation}
\Lambda = 
 \begin{pmatrix}
  k^2 n_3^2 -\omega^2 -\eta \omega kn_1t_2 \, \, \, & \eta \omega k t_1n_1 & \eta\omega k t_2 n_3 \\
 0 & k^2-\omega^2&  -\eta\omega k t_1/n_3 \\
-(k^2n_1+\eta\omega k t_2)n_3 \, \, & \eta\omega k t_1n_3 & -(\omega^2 +\eta\omega k t_2 n_1)
 \end{pmatrix}. \nonumber
 \label{eq:Lambda}
\end{equation}
The dispersion relation is obtained from the equation 
\begin{equation}
\begin{aligned}
\det \Lambda & = (k^2-\omega^2)[k^2 n_3^2(\eta^2t_2^2 -1) \\
& +(\omega + \eta k n_1t_2)^2 +\eta^2 k^2 t_1^2 ]\omega^2 =0,
\end{aligned}
\end{equation}
which can be written more conveniently in terms of the phase velocity $v=\omega/k$ as
\begin{equation}
 (1-v^2)[(v+ \eta n_1 t_2)^2 - n_3^2(1-\eta^2)] v^2 =0. \label{eq:phase_v}
\end{equation}

\section{Low frequency limit in a cold magnetized plasma}
\label{app:low_omega}
For completeness, we rederive the dispersion equation for waves in cold, magnetized plasma,
 and show that FFE emerges in the low frequency and high magnetization limit.

As in appendix \ref{app:waves}, we denote the linear wave fields as $\pmb{e}$ and $\pmb{b}$, 
but now choose, for convenience,  $\pmb{B}_0 = B_0\hat{z}$ and $\pmb{k}= k(\sin\theta \hat{x} + \cos\theta \hat{z})$.
To first order Eq. (\ref{eq:kinetic}) reads:
\begin{equation}
    \partial_t \pmb{v}_\pm  = \pm \frac{e}{m_\pm}(\pmb{e} + \pmb{v}_\pm \times \pmb{B}_0).
\end{equation}
For periodic waves it can be readily solved to yield the electric current,
\begin{equation}
    \pmb{J} = \stackrel{\leftrightarrow}{\sigma} \pmb{e},
\end{equation}
expressed here in terms of the conductivity tensor 
\begin{equation}
\stackrel{\leftrightarrow}{\sigma} = \frac{i\omega}{4\pi}
\begin{pmatrix}
 S -1 & iD & 0 \\
 -iD & S-1 & 0 \\
 0 & 0& P -1
\end{pmatrix},
\end{equation}
%
%
with 
\begin{equation}
\begin{aligned}
S &= 1 - \frac{\omega_{p+}^2}{\omega^2- \omega_{c+}^2}- \frac{\omega_{p-}^2}{\omega^2- \omega_{c-}^2},\\
D & = \frac{\omega_{p+}^2 \omega_{c+}}{\omega(\omega^2-\omega_{c+}^2)} - \frac{\omega_{p-}^2 \omega_{c-}}{\omega(\omega^2-\omega_{c-}^2)} ,\\
P &=  1 - \frac{\omega_{p+}^2}{\omega^2}- \frac{\omega_{p-}^2}{\omega^2}.
\end{aligned}
\end{equation}
Here $\omega_{p\pm} = (4\pi e^2 n_\pm/m_\pm)^{1/2}$ is the plasma frequency and $\omega_{c\pm} = eB_0'/m_\pm$ the gyro frequency
of species +(-).
Substituting into Maxwell's equation, one finds:
\begin{equation}
 \pmb{k}\times(\pmb{k}\times\pmb{e}) + \omega^2 \left(I - \frac{4\pi i}{\omega}\stackrel{\leftrightarrow}{\sigma}\right) \pmb{e}=0. 
 \label{eq:App_B_Max}
\end{equation}
We denote the angle between $\pmb{k}$ and $\pmb{B}_0$ by $\theta$ and introduce the refractive index $n=k/\omega$.
Then, Eq. ( \ref{eq:App_B_Max}) can be written as $\Lambda_{ij} e_j =0$, where 
\begin{equation}
\Lambda = 
\begin{pmatrix}
 S -n^2 \cos^2\theta& -iD & n^2 \cos\theta\sin\theta \\
 iD & S- n^2 & 0 \\
n^2 \cos\theta\sin\theta & 0& P - n^2\sin^2\theta
\end{pmatrix}.
\label{eq:Lambda_coldB}
\end{equation}
%
The dispersion relations are now obtained from $\det{\Lambda}=0$:
\begin{equation}
\begin{aligned}
 (S - n^2)[SP & -n^2(P\cos^2\theta + S \sin^2\theta)] \\
 & -D^2(P -n^2\sin^2\theta) =0 .  
 \end{aligned}
\end{equation}

In the low frequency limit, $\omega \ll \omega_{p\pm}, \omega_{c\pm}$, we find the following expansion, expressed in terms of the 
magnetizations  $\sigma_{\pm}= (\omega_{c\pm}/\omega_{p\pm})^2 = B_0^2/4\pi m_\pm n_\pm$:
\begin{equation}
\begin{aligned}
S & \to  1 + \frac{1}{\sigma_+} + \frac{1}{\sigma_-},\\
D & \to  \frac{1}{\sigma_-}\frac{\omega}{\omega_{c-}} - \frac{1}{\sigma_+}\frac{\omega}{\omega_{c+}} ,\\
P & \to   - \frac{\omega_{p+}^2}{\omega^2}- \frac{\omega_{p-}^2}{\omega^2}.
\label{eq:app_low_omega}
\end{aligned}
\end{equation}
 Thus, to leading order $D=0$ and, neglecting terms of order $S/P$, the dispersion equation reduces to \footnote{The low frequency 
 limit holds for $|P|\gg |S| \tan^2\theta$.}
\begin{equation}
    (S-n^2)(S-n^2\cos^2\theta)=0, 
\end{equation}
which is equivalent to Eq. (\ref{eq:phase_v}) for $\eta=0$.  For $\theta\ne \pi/2$ the first term yields the two fast modes and the second term
two Alfv\'en modes. Note that $\sqrt{1/S} = \sqrt{\sigma/(\sigma+1)}$,
here $\sigma^{-1} = \sigma_{+}^{-1} + \sigma_-^{-1}$ is the total magnetization, is simply the Alfv\'en speed $v_A$ of the cold plasma,
which in the force-free limit, $\sigma\to\infty$, approaches 1.  For the fast modes Eq. \ref{eq:Lambda_coldB} yields the polarization
$\pmb{e} = (0,1,0)$, and for the Alfv\'en modes $\pmb{e}=(\sqrt{1-S^2\tan^2\theta/P^2},0, S \tan\theta/P) \xrightarrow[\omega \to 0]{} (1,0,0)$, 
so $\pmb{e}\cdot\pmb{B}=0$ in this limit, as required.

It is seen that despite the fact that the component of the wave electric field along the background magnetic field, $e_3$,
is vanishingly small, the electric current is actually directed along $\pmb{B}_0$. This is because the parallel conductivity 
tends to infinity in this limit.  More precisely, the leading order of the conductivity tensor is
\begin{equation}
\stackrel{\leftrightarrow}{\sigma} = \frac{i\omega}{4\pi}
\begin{pmatrix}
 1/\sigma & 0 & 0 \\
0 & 1/\sigma & 0 \\
 0 & 0&  P-1 
\end{pmatrix},
\label{eq:conductivity_low_omega}
\end{equation}
and the electric current,
\begin{equation}
\begin{aligned}
 \pmb{J} & = (i\omega/4\pi)(\sigma^{-1},0,  \tan\theta) \\
 & \xrightarrow[\sigma \to \infty]{}
 (i\omega/4\pi)(0,0,\tan\theta),
 \label{eq:J_app_B}
 \end{aligned}
\end{equation}
in agreement with the result of appendix \ref{app:waves}.

\section{Vlasov equation}
\label{app:vlasov}
The collisionless Vlasov equation for species $s$, having mass $m_s$ and electric charge $q_s$, 
can be expressed as
\begin{equation}
    p^\mu \frac{\partial f_s}{\partial x^\mu} + Q_s^i \frac{\partial f_s}{\partial p^i}=0,
\end{equation}
where $f_s(x,p)$ denotes the phase space density (distribution function), and 
$Q_s^i = (q_s/m_s) F^{i\nu}p_\nu$ the force acting on the particles.  

The mean properties of the system can be described by the leading order moments of the distribution function. 
The first and second moments correspond, respectively,  to the particle flux
and the energy-momentum tensor:
\begin{equation}
\begin{aligned}
    N_s^\mu & =\int p^\mu f_s \frac{d^3p}{p^0},\label{eq:app_C_flux}\\
    T_s^{\mu\nu} & = \int p^\mu p^\nu f_s \frac{d^3p}{p^0}.
\end{aligned}
\end{equation}
From the Vlasov equation  a set of equations for the moments can be derived.  Formally, this set is infinite. 
However, it can be truncated by introducing a closure condition.
One usually posits that there  exist a local frame in which the distribution function is isotropic. This defines the 
proper density and pressure (see, e.g., \citealt{lev2012}).  In addition, an equation of state is needed.  

Let $\chi(p)$ represent a generic moment of the Vlasov
equation:
\begin{equation}
   \int \chi(p) \left\{p^\mu \frac{\partial f_s}{\partial x^\mu} + Q_s^i \frac{\partial f_s}{\partial p^i}\right\}\frac{d^3p}{p^0}=0.
   \label{eq:moment_generic}
\end{equation}
Integrating by parts the second term, noting that $\partial Q_s^i/\partial p^i=0$ and using $\chi Q_s^i \partial f_s/\partial p^i = \partial (\chi Q_s^if_s)/\partial p^i - Q_s^i f_s \partial \chi/\partial p^i$, we obtain
\begin{equation}
   \int \chi(p) Q_s^i \frac{\partial f_s}{\partial p^i} \frac{d^3p}{p^0}=  -\int Q_s^i f_s  \frac{\partial \chi(p)}{\partial p^i} \frac{d^3p}{p^0}.
\end{equation}
Substituting into Eq. (\ref{eq:moment_generic}) one finds:
\begin{equation}
 \frac{\partial}{\partial x^\mu} \int \chi(p) p^\mu f_s \frac{d^3p}{p^0} = \int Q_s^i f_s  \frac{\partial \chi(p)}{\partial p^i} \frac{d^3p}{p^0}.
\end{equation}
In particular, $\chi(p)=1$ yields conservation of particle flux,
\begin{equation}
   \partial_\mu N^\mu  = 0,
\end{equation}
and $\chi(p)= p^\mu$ the equation for the stress tensor:
\begin{equation}
    \partial_\mu T_s^{\mu\nu}  = q_s F^{\nu\mu}N_{s\mu}.
\end{equation}
For $s=\pm$ we obtain Eqs.(\ref{eq:continuity}),(\ref{eq:T=FJ}).

The Vlasov equation can be solved using the method of characteristics \citep[e.g.,][]{rendall199}.  In components they are 
\begin{equation}
\begin{aligned}
  d X_s^k/dt & = P_s^k/P_s^0,\\
  d P_s^k/dt & =  Q_s^k/P_s^0 = q_s(E^k + \epsilon^{kij}V_{si}B_j),
  \label{eq:characteristica}
\end{aligned}
\end{equation}
here $V_s^i =P_s^i/P_s^0$ is the 3-velocity.  The general solution can be written  as
\begin{equation}
  f_s(t,x^i,p^j) = g_s[X_s^k(0,x^i,p^j),P_s^l(0,x^i,p^j)],
  \label{eq:initial_data}
\end{equation}
where $g(x^i,p^j) = f(0,x^i,p^j)$ is the initial data (i.e., the state of the system at $t=0$), assumed to be given, and $X^k(t,x^i,p^j),P^l(t,x^i,p^j)$ 
is the solution of Eqs. (\ref{eq:characteristica}) that satisfies $X^k(t,x^i,p^j)=x^k$, $P^l(t,x^i,p^j)=p^l$ .  
The electric 4-current is give by 
\begin{equation}
  J^\mu = \sum_s q_sN^\mu_s = \sum_s \left \{ q_s \int p^\mu g_s \frac{d^3p}{p^0}\right\},  
  \label{eq:appC_J}
\end{equation}
with $g_s$ defined in Eq. (\ref{eq:initial_data}).  Eqs. (\ref{eq:characteristica}),(\ref{eq:appC_J}) together with Maxwell's equations 
defines a closed set that can be solved. 

As an example, consider a uniform electric zone, $B^i=0$, with a two species neutral plasma, $q_+= +e, q_- = -e$.
Suppose that initially all particles are at rest (zero temperature).  Then $g_s(x^i,p^j) = N_0 p^0 \delta^3(\pmb{p})$, 
where $N_0 =N_+ = N_-$ is the density of each species.  From Eq. (\ref{eq:appC_J}) we find 
\begin{equation}
 \pmb{J} = eN_0( \pmb{V}_+ - \pmb{V}_-) 
  \label{eq:appC_J_example}
\end{equation}
and from eq (\ref{eq:characteristica}), after dividing by the mass $m_\pm$, 
\begin{equation}
 d\pmb{U}_\pm/dt = \frac{e}{m_\pm}\pmb{E}.
\end{equation}
Together with Maxwell's equation, $\partial_t \pmb{E} = - 4\pi \pmb{J}$, we obtain the set (\ref{eq:ME4})-(\ref{eq:kineticE}).
Thus, this is equivalent to the two-fluid description. 

\section{Linear waves around oscillating electric field}
\label{app:linear_waves}

Consider initial state with constant electric field $\pmb{E}_0'$ and neutral plasma at rest with density $n_+(0)=n_-(0)=N_0$.
Define 
\begin{equation}
\begin{aligned}
\pmb{E}(t,z) &=  \pmb{E}(t) +  \delta \pmb{E}(t,z),\\
\pmb{B}(t,z) & =   \delta \pmb{B}(t,z),\\
\pmb{u}_\pm(t,z) & =  \pmb{u}_{0\pm}(t) +  \delta \pmb{u}_\pm(t,z),\\
n_\pm(t,z) &= n_\pm(t) +\delta n_\pm(t,z)
\end{aligned}
\end{equation}
where $\delta \pmb{E}(t,z), \delta \pmb{B}(t,z), \delta \pmb{u}_\pm(t,z), \delta n_\pm(t,z)$ denote linear perturbations
around the (time dependent) background solution $n_\pm(t), \pmb{u}_\pm(t), \pmb{E}(t)$, that can be large.
The electric current can be expanded as $J^\mu(t) = e(n_+ u_+^\mu - n_- u_-^\mu)$,
 $\delta J^\mu (t) = e(\delta n_+ u^\mu_+ + n_+ \delta u^\mu_+ - \delta n_- u^\mu_-  - n_- \delta u^\mu_-)$.
 %
 %
The lowest order equations for $\pmb{E}(t), \pmb{u}_\pm(t), n_\pm(t)$ reduce to Eqs (\ref{eq:ME4})-(\ref{eq:kineticE}).  The solution
yields the background plasma oscillations.  The next order yields the equations for the superposed linear waves:
\begin{equation}
\begin{aligned}
\nabla\cdot \delta\pmb{E} &= \delta \rho_e,\\
\nabla\cdot \delta\pmb{B} &= 0,\\
\partial_t \delta\pmb{B} &= -\nabla\times \delta \pmb{E},\\
\partial_t \delta \pmb{E} &=\nabla\times \delta\pmb{B} - 4\pi \delta \pmb{J},\\
\partial_t \delta \pmb{u}_\pm &= \pm \frac{e}{m_\pm}[ \delta \pmb{E} +\pmb{u}_\pm(t) \times \delta\pmb{B}],\\
\partial_t (\delta n_\pm \gamma_\pm +n_\pm \delta \gamma_\pm) &= 
(\pmb{u}_\pm(t)\cdot\nabla)\delta n_\pm +n_\pm(t)\nabla\cdot\delta\pmb{u}_\pm,
\end{aligned}
\end{equation}
that can be solved once $n_\pm(t)$ and $\pmb{u}_\pm(t)$ are computed.   

%

\end{document}